# DESIGN AND ANALYSIS OF A NOVEL DIGITAL IMAGE ENCRYPTION SCHEME

Narendra K Pareek

University Computer Centre, Vigyan Bhawan,
M L Sukhadia University, Udaipur-313001, India.
npareek@yahoo.com

## ABSTRACT

*In this paper, a new image encryption scheme using a secret key of 144-bits is proposed. In the substitution process of the scheme, image is divided into blocks and subsequently into color components. Each color component is modified by performing bitwise operation which depends on secret key as well as a few most significant bits of its previous and next color component. Three rounds are taken to complete substitution process. To make cipher more robust, a feedback mechanism is also applied by modifying used secret key after encrypting each block. Further, resultant image is partitioned into several key based dynamic sub-images. Each sub-image passes through the scrambling process where pixels of sub-image are reshuffled within itself by using a generated magic square matrix. Five rounds are taken for scrambling process. The propose scheme is simple, fast and sensitive to the secret key. Due to high order of substitution and permutation, common attacks like linear and differential cryptanalysis are infeasible. The experimental results show that the proposed encryption technique is efficient and has high security features.*

## 1. INTRODUCTION

With the rapid growth of computer networks and advances in information technology, a huge amount of digital data is being exchanged over unsecured channels. Major part of transmitted information, either confidential or private, demands for security mechanisms to provide required protection. Therefore, security has become an important issue during the storing and transmission of digital data. The security of images is an application layer technology to guard the transmitted information against unwanted disclosure as well as to protect the data from modification while in transit. It includes several aspects like copyright protection, authentication, confidentiality and access control. It is argued that traditional encryption algorithms are not preferable for image encryption for two reasons. One is that the image size is always more than text size. Therefore, traditional cryptosystems take longer time to encrypt the image data. The other is that the size of decrypted text must be equal to the original text size. However, this requirement is not necessary for image data. Due to the characteristic of human perception, a small distortion in decrypted image is usually acceptable. Three different ways to protect digital data from unauthorized eavesdropping are cryptography, steganography and watermarking. Among these three techniques, cryptography has become one of the major tools to provide high level of security. Cryptography deals with the development of techniques for converting information between intelligible and unintelligible forms. It deals with the content confidentiality and access control. In secure communications using cryptography, which is the main focus of the present work, the encryption and decryption operations are guided by one or more keys. Techniques that use same secret key for encryption and decryption are grouped under private key cryptography. Alternately, encryption and decryption keys are different or computationally it may not be feasible to derive one key even though the knowledge of other key, such cryptographic methods are known as public key cryptography. The present work focuses on the development of private key image cryptographic algorithm for providing high level of security.

## 1.1 Related work

In order to protect digital images from unauthorized users doing illegal reproduction and modifications, a variety of image encryption schemes have been proposed. The various ideas used in the existing image encryption techniques can be classified into three major categories: position permutation [1,2,7,10], value transformation [3,4,8,12] and the combination form [5,6,9,11,13,14]. The position permutation algorithms scramble the data position within the image itself and usually have low security. On the other hand, the value transformation algorithms transform the data value of the original signal and have the potential of low computational complexity and low hardware cost. Finally, the combination forms perform both position permutation and value transformation and usually have the potential of high security. In recent years, a number of different image encryption schemes have been proposed in order to overcome image encryption problems. A few image encryption techniques suggested recently are discussed in the following paragraph in brief.

In 2010, Yoon and Kim [7] developed a chaotic image cipher in which initially a small matrix was generated using chaotic logistic map. Authors constructed a large permutation matrix from generated small matrixes. The constructed permutation matrix was used to permute plain image pixels. Further, a new chaotic image cipher was suggested by Ismail et al [8] in which they used two chaotic logistic maps as well as an external secret key of 104-bits size. Control parameters for both chaotic logistic maps were generated from the used external secret key. They also employed a feedback mechanism in their image cipher to make system more secured. In 2011, Jolfaei and Mirghadri [9] suggested a chaotic image cipher based on pixel shuffling, using baker map, and modified version of simplified AES (S-AES), developed by Musa et al. in 2003 [15]. Further, Nayak et al. [10] proposed a chaotic image cipher using logistic map. In the algorithm, permutation of image pixels was made on the basis of index position of generated chaotic sequence. Sathishkumar and Bagan [11] suggested a chaotic image cipher based in which block permutation, pixel permutation and value transformation were used. Next, chaotic image cipher using two chaotic logistic maps and a secret key of 80-bits was suggested by Chen and Chang [12]. Indrakanti and Avadhani [13] developed a non-chaotic image cipher in which encryption was achieved in three processes. In the last process, a key was generated which keep all information about encryption process. Next, Xin et al [20] proposed a chaos based cryptosystem based on diffusion and confusion. To improve the security of the confusion module, authors introduced diffusion mechanism in confusion stage through a lightweight bit level permutation algorithm. Peng et al [21] proposed an image block encryption algorithm based on three-dimensional Chen chaotic dynamical system. The algorithm used a secret key of 192-bit size with an image block of 32-bit long. In the algorithm, Chen's system generated a chaotic sequence that was inputted to a designed function G which was iterated several times. Further, Sathishkumar et al [22] suggested an image encryption scheme using multiple chaotic based circular mapping. Based on the initial conditions, each map produces random numbers from orbits of the maps. Among those random numbers, a particular number from a particular orbit was selected as a key for the encryption algorithm. Further, input image was reshuffled and divided into various sub-blocks. Then the position permutation and value permutation was applied to each binary matrix based on multiple chaos maps. Jamei et al [23] proposed an image encryption using chaotic signals and complete binary tree. In the method, perfect binary tree was utilized to increase complexity in the encryption algorithm. Next, Sathyanarayana et al [24] designed an encryption algorithm focusing on the application of properties of finite fields and elliptic curves. Additive and Affine encryption schemes using six schemes of key sequences obtained from random elliptic curve points were designed. In the developed algorithm, cyclic elliptic curves of the form $y^2+xy =x +ax^2+b$, $a$, $b \in GF(2^m)$, $b \neq 0$ with order M was considered in the design of a symmetric key image encryption scheme. Pareek et al [25] proposed a lossless digital image encryption scheme based on the permutation and substitution architecture. In the algorithm, plain image was divided into squared sub-images and then reshuffled them. Further, in the permutation process a simple arithmetic, mainly sorting and differencing, was performed to achieve the substitution. Hybrid model of cellular automata and chaotic signal was proposed for image encryption by Fateri and Enayatifar [26]. In this method, 8-bits mask was used for changing the pixel gray level of main image. For changing each pixel gray level, value of each bit of the mask was selected by one of the 256 cellular automat standard rules. One of the 256 cellular automat standard rules was determined by chaotic signal.

## 1.2 Paper outline

In available literature, most of suggested image encryption schemes are based on chaotic dynamical system. In this paper, a non-chaos based image encryption scheme using an external key of 144-bits is presented. The proposed encryption scheme uses both pixel substitution as well as pixel permutation process. A feedback mechanism, to avoid differential attack and make the cryptosystem more robust, is also applied. The proposed encryption scheme has high encryption rate, requires less computation and sensitive to small changes in the secret key so even with the knowledge of the approximate key values, there is no possibility for the attacker to break the cipher. The proposed scheme is simple, fast and secured against any attack.

The rest of the paper is organized in the following manner. In section 2, various processes used in the proposed algorithm as well as detail of encryption algorithm is discussed. Simulation results and security analysis are provided in Section 3. Finally, the conclusions are drawn in Section 4.

## 2. PROPOSED ENCRYPTION ALGORITHM

Image data have strong correlations among adjacent pixels forming intelligible information. To encrypt the image, this intelligible information needs to be reduced by decreasing the correlation among the adjacent pixels. The proposed encryption algorithm does this by modifying the pixel values of the image as well as reshuffling the pixels of the resultant image within itself. In following paragraph, functions of various processes used in the proposed algorithm are discussed.

### 2.1 Key mixing process

Two kinds of key mixing processes, called FKM and BKM, are used in the proposed algorithm. In both the processes, image block is divided into sub-blocks ($p_1, p_2..p_{18}$) and each sub-block ($p_i$) is modified by using sub-key ($k_i$), its previous sub-block ($p_{i-1}$) and sub-block ($p_i$) itself. A similar process is used in the BKM process.

Forward key mixing (FKM) process
    i=2
    temp=$p_1$, $p_1$= $p_1 \oplus k_1$
    if (i<=18) then temp1=$p_i$ , $p_i = p_i \oplus k_i \oplus$ temp, temp=temp1
    else   i=i+1
    endif

Backward key mixing (BKM) process
    i=17
    temp=$p_{18}$, $p_{18}$= $p_{18} \oplus k_{18}$

if (i >=1) then temp1=$p_i$ , $p_i = p_i \oplus k_i \oplus$ temp, temp=temp1
else   i=i-1
endif

## 2.2 Substitution process

In this process, bitwise operations are performed on pixels of sub-blocks to change their properties. Pictorial view of substitution process used in the proposed system is shown in the Figure (1) and its equivalent description is as follows:

1. Take a plain image block and divided it into eighteen equal sub-blocks named as $p_1, p_2, p_3, \ldots, p_{18}$.
2. set position=5
3. set i=1, $T_i = p_i$
4. if (position=5) then $p_i = p_i \oplus (k_j \mod 18)$
5. bv=$T_i$ >> position
6. set i=i+1, $T_i = p_i$
7. perform operation on $p_i$ as shown in Table (1)
8. if (i<=18) then go to step (5)
9. if (position=7) then
    for (i=1;i<=18;i=i+1)  $k_i = k_i \oplus p_i$
    j=j+1
    stop
   endif
10. if (position=5) then perform BKM process
    else perform FKM process
    endif
11. position=position+1
12. go to step (3)

In the substitution process, image block to be modified is partitioned into eighteen equal sub-blocks ($p_1, p_2, \ldots p_{18}$) having 8-bits ($b_1, b_2, \ldots b_8$) each. Further, each sub-block passes through the following three steps:

*First step*: In this step, first sub-block ($p_1$) is modified by xoring it with first sub-key ($k_1$) and remaining sub-blocks ($p_2 \ldots p_{18}$) are modified by applying the operation as shown in the Table (1). The operation, to be applied on each sub-block ($p_i$), depends on first three most significant bits ($b_1 b_2 b_3$) of their previous sub-block ($p_{i-1}$). For example, value of first two sub-blocks $p_1$ and $p_2$ are 203 (**110**01011) and 185 (10111001) respectively. The value of sub-block ($p_2$), after applying operation on it corresponding to bits value 6 (110) of its previous sub-block ($p_1$), change to 141 (10001101). Further, BKM process, as described in Sub section 2.1, is applied on the resultant block.

*Second step*: Resultant block received from the first step is processed again. In this step, first sub-block ($p_1$) remains unchanged and remaining sub-blocks ($p_2 \ldots p_{18}$) are modified by applying the operation as shown in the Table (1). The operation, to be applied on each sub-block ($p_i$), depends on first two most significant bits ($b_1 b_2$) of their previous sub-block ($p_{i-1}$). For example, value of first two sub-blocks $p_1$ and $p_2$ are 139 (**10**001011) and 185 (10111001) respectively. The value of second sub-block ($p_2$), after applying the operation on it corresponding to bits value 2 (10) of its previous sub-block ($p_1$), change to 70 (01000110). Further, FKM process, as described in Sub section 2.1, is applied on the resultant block and passes it to third step for further processing.

*Third step*: In this step, first sub-block ($p_1$) remains unchanged and remaining sub-blocks ($p_2...p_{18}$) are modified by applying the operation as shown in the Table 1. The operation, to be applied on each sub-block ($p_i$), depends on left most significant bits ($b_1$) of their previous sub-block ($p_{i-1}$). For example, value of first two sub-blocks $p_1$ and $p_2$ are 139 (**1**0001011) and 185 (10111001) respectively. The value of second sub-block ($p_2$), after applying the operation on it corresponding to bits value 1 of its previous sub-block ($p_1$), change to 92 (01011100). Further, FKM process, as described in Sub section 2.1, is applied on the resultant block.

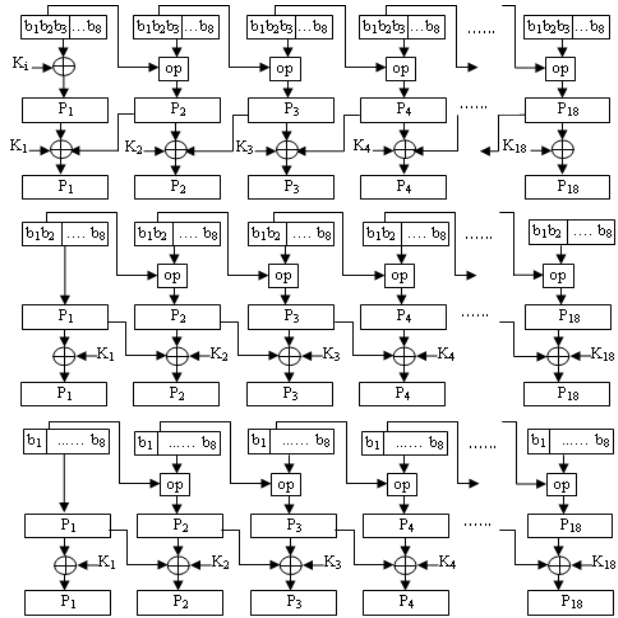

Figure 1. Architecture of substitution process.

Table 1. Bits value and their corresponding operations.

| Bits value | Operation on sub-block for encryption |
|---|---|
| 0 | Circular left shift by one position i.e. $p_i = p_i <<1$ |
| 1 | Circular right shift by one position i.e. $p_i = p_i >>1$ |
| 2 | Invert all bits i.e. $p_i$ = Not ($p_i$) |
| 3 | $p_i = p_i \oplus (k_i \bmod 18)$ |
| 4 | $p_i$ = Not ($p_i$) $\oplus$ ($k_i$ mod 18) |
| 5 | $p_i$ = Not ($p_i \oplus (k_i \bmod 18)$) |
| 6 | $p_i$ = Not (circular left shift by one position ) |
| 7 | $p_i$ = Not (circular right shift by one position ) |

In the proposed system, the secret key is modified from the previously encrypted plain image block for processing the next block. This feedback mechanism, avoid differential attack and make the system robust.

### 2.3 Permutation process

In this process, pixel's positions of sub-image are scrambled within itself. Below, the procedure to scramble the pixel's position by applying the following pixel swapping function to sub-image is shown. Pixel swapping function requires sub-image dimension as well as sub-image as its argument.

```
Function pixelswap(integer N, integer SubImage[N][N])
  begin
    set  Nsqr=N*N, index=1, Temp[Nsqr]
    for (row=1; row<=N; row=row+1)
  {  for(col=1; col<=N; col=col+1)
     { Temp[index]=SubImage[row][col]
       SubImage[row][col]=0
       index=index+1 }}
   set row=1, col=N div 2 +1
   for (index=1;  index<=Nsqr;  index=index+1)
  { SubImage[row][col]= Temp[index]
    row=row-1, col=col+1
   if ((row=0 and col>N) or (SubImage[row][col]#0)) then
       {row=row+2, col=col-1}
   else if (row=0) then row=N
   else if (col>N) then col=1}
 end
```

In the following example, how positions of pixels are reshuffled within sub-image is shown. Pixels of sub-image of dimension 5x5 are shown in Table 2(a). Further, a magic square of same dimension is constructed as shown in Table 2(b). Now, permute pixel of sub-image referencing constructed magic square. The resultant sub-image pixels are shown in the Table 2(c).

Table 2. Pixel of a sub-image, magic square matrix and pixel of resultant sub-image is shown in (a), (b) and (c) respectively.

| 12 | 242 | 130 | 55 | 99 |
|---|---|---|---|---|
| 167 | 203 | 57 | 17 | 226 |
| 65 | 69 | 6 | 77 | 133 |
| 219 | 245 | 101 | 45 | 77 |
| 3 | 18 | 55 | 201 | 55 |

(a)

| 17 | 24 | 1 | 8 | 15 |
|---|---|---|---|---|
| 23 | 5 | 7 | 14 | 16 |
| 4 | 6 | 13 | 20 | 22 |
| 10 | 12 | 19 | 21 | 3 |
| 11 | 18 | 25 | 2 | 9 |

(b)

| 245 | 201 | 12 | 57 | 133 |
|---|---|---|---|---|
| 55 | 99 | 203 | 77 | 219 |
| 55 | 167 | 6 | 77 | 18 |
| 226 | 69 | 45 | 3 | 130 |
| 65 | 101 | 55 | 243 | 17 |

(c)

## 2.4 The Complete encryption algorithm

Proposed algorithm has two major processes - pixels substitution and pixels permutation. Both substitution and permutation processes are completely secret key dependent. A feedback mechanism is applied in substitution process to avoid differential attack. Steps of algorithm are discussed below.

1) The proposed encryption scheme uses a secret key of 144-bits size. The secret key is divided into eighteen equal parts of 8-bits each referred as sub-keys.
$$k=k_1k_2k_3 \ldots\ldots\ldots\ldots k_{18}$$ ……. (1)
2) Plain image is divided into several blocks comprising eighteen bytes each.
3) Each image block is passed through the substitution process. When a processing of a block is over, the key is modified from its previously encrypted image block. Resultant pixels are written in a file.
4) set iteration=5
5) Take original secret key and generate key dependent odd integer random number in the range 50 to 200, say N, using any random number generator.
6) Partial encrypted image is divided into several non-overlapping squared sub-images each of size NxN.
7) Take sub-images one by one and passes through the permutation process.
8) Decrease rounds value by one and repeat the above steps from (5) till iteration value is non-zero. Finally, the encrypted pixels of the sub-image are written in a file.

## 3. PERFORMANCE AND SECURITY ANALYSIS

Several tests on USC-SIPI image database (freely available at http://sipi.usc.edu/database/) to justify the security and performance of the proposed image cipher have been done. An ideal image cipher should resist against all kinds of known attacks such as cryptanalytic, statistical and brute force attacks. In this section, security analysis of the proposed image encryption scheme such as statistical analysis, key space analysis, sensitivity analysis etc to prove that the proposed image cipher is effective and secure against the most common attacks is discussed. The proposed algorithm has been implemented in C programming language. For the analysis of image data, Matlab application tool was used.

### 3.1 Statistical analysis

In the literature, it was found that most of the existing image ciphers have been successfully cryptanalyzed with the help of statistical analysis. To prove the robustness of the proposed encryption scheme, statistical analysis has been performed which demonstrates its superior confusion and diffusion properties results in a strongly resisting nature against the statistical attacks. This is done by testing the distribution of pixels of the cipher images, study of correlation among the adjacent pixels in the cipher image, information entropy and the correlation between the plain and cipher images.

#### 3.1.1 Distribution of pixels

Histograms of several cipher images and their corresponding plain images having widely different contents and sizes have been analysed. One example of histogram analysis for well known image '*Lena*' is shown in Figure 2. Histograms of red, blue and green components of image (Figure 2(a)) are shown in Frames (b), (c) and (d) respectively. In Frames (f), (g) and (h) respectively, the histograms of red, blue and green components of the cipher image (Figure 2(e)) are shown. Comparing the histograms, it is found that encryption process returns noisy images. Histograms of cipher images, approximated by uniform distribution, are quite different from that of the plain image and contain no statistical resemblance to the plain image. This is consistent with the perfect security defined by Shannon [16] and the proposed encryption scheme resists against the known-plaintext attack.

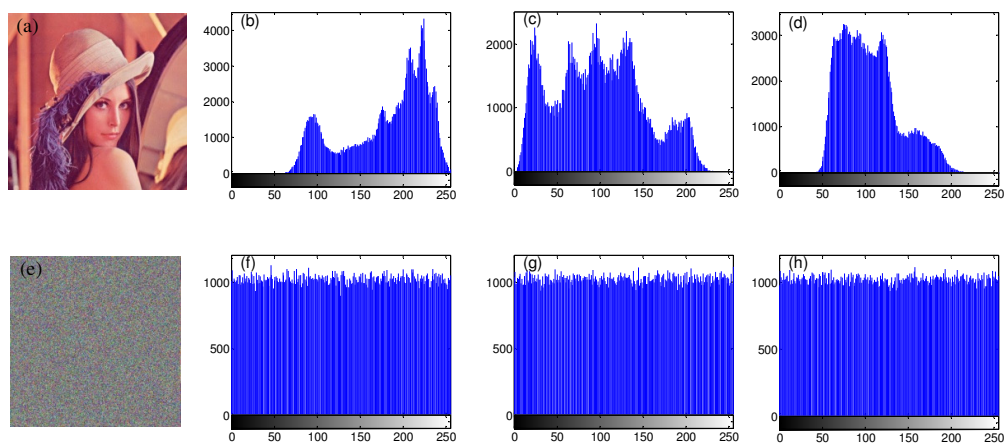

Figure 2. Histograms corresponding to RGB components of plain image *'Lena'* and its corresponding cipher image.

### 3.1.2 Correlation between plain and cipher images

An extensive study of the correlation between pairs of plain image and their corresponding cipher image produced using the proposed encryption scheme by computing correlation coefficient (CR) between RGB components of the plain images and corresponding cipher images have been done. Results for a few images are shown in Table 3. Since the correlation coefficients shown in the Table 3 are very small (C≈0), it indicates that the plain images and their corresponding cipher images are completely independent of each other.

Table 3. CR between plain images and their corresponding cipher images.

| Image size | $C_{RR}$ | $C_{RB}$ | $C_{RG}$ | $C_{GR}$ | $C_{GG}$ | $C_{GB}$ | $C_{BR}$ | $C_{BG}$ | $C_{BB}$ |
|---|---|---|---|---|---|---|---|---|---|
| 200 x 200 | 0.0074 | 0.0076 | 0.0031 | 0.0060 | 0.0072 | 0.0032 | 0.0048 | 0.0050 | 0.0026 |
| 200 x 200 | -0.0083 | -0.0012 | -0.0079 | -0.0062 | 0.0001 | -0.0091 | -0.0026 | 0.0066 | -0.0009 |
| 512 x 512 | -0.0041 | 0.0013 | 0.0019 | -0.0031 | -0.0003 | 0.0025 | -0.0021 | -0.0016 | 0.0021 |
| 512 x 768 | -0.0017 | 0.0066 | 0.0062 | -0.0009 | 0.0045 | 0.0046 | -0.0026 | 0.0006 | 0.0009 |
| 600 x 326 | 0.0004 | 0.0021 | 0.0016 | -0.0010 | 0.0015 | 0.0028 | -0.0014 | -0.0010 | 0.0009 |
| 640 x 480 | -0.0002 | -0.0004 | 0.0028 | 0.0005 | -0.0004 | 0.0028 | 0.0004 | -0.0005 | 0.0026 |
| 668 x 480 | -0.0009 | -0.0014 | 0.0010 | -0.0002 | -0.0012 | 0.0013 | 0.0007 | -0.0007 | 0.0018 |
| 800 x 600 | -0.0026 | 0.0024 | 0.0016 | -0.0018 | 0.0030 | 0.0013 | -0.0011 | 0.0026 | 0.0008 |
| 900 x 600 | 0.0027 | -0.0007 | 0.0000 | 0.0030 | -0.0003 | 0.0003 | 0.0026 | -0.0006 | -0.0006 |

### 3.1.3 Correlation analysis of adjacent pixels

Correlations between two vertically and horizontally adjacent pixels in various plain images and their corresponding cipher images have been analysed. In Figure 3, the distributions of horizontally adjacent pixels of red, green and blue components in the image '*Lena*' and their corresponding cipher image is shown. Particularly, in Frames (a), (b) and (c), depict the distributions of two horizontally adjacent pixels of red, green and blue components respectively in the plain image (Figure 2(a)). Similarly in Frames (d), (e) and (f) respectively, the distributions of two horizontally adjacent pixels in its corresponding cipher image (Figure 2(e)) have been depicted. Similarly, in Figure 4, the distributions of vertically adjacent pixels of red, green and blue components in the plain image '*Lena*' and its corresponding cipher image is shown. It is observed from correlation charts and Table 4 that there is a negligible correlation between the two adjacent pixels in the cipher image. However, the two adjacent pixels in the plain image are strongly correlated. Correlation in the cipher images is very small or negligible when the proposed encryption scheme is used. Hence the proposed scheme has good permutation and substitution properties.

Table 4. CR for two adjacent pixels in the plain and its cipher image.

| | | Correlation coefficient between adjacent pixels | | |
|---|---|---|---|---|
| | | Red | Green | Blue |
| Horizontal | Plain image | 0.9608 | 0.9845 | 0.9850 |
| | Cipher Image | 0.0489 | -0.0624 | -0.0666 |
| Vertical | Plain image | 0.7778 | 0.8996 | 0.9731 |
| | Cipher image | -0.0031 | 0.0094 | 0.0081 |

### 3.1.4 Information entropy

Illegibility and indeterminateness are the main goals of image encryption. This indeterminateness can be reflected by one of the most commonly used theoretical measure -

International Journal of Network Security & Its Applications (IJNSA), Vol.4, No.2, March 2012information entropy. Information entropy expresses the degree of uncertainties in the system and defines as follow.

$$H(m) = -\sum_{i=0}^{2^N-1} P(m_i) \log_2[P(m_i)] \qquad \ldots\ldots (2)$$

where $P(m_i)$ is the emergence probability of $m_i$. If every symbol has an equal probability, i.e., $m=\{m_0, m_1, m_2, \ldots m_{2^8-1}\}$ and $P(m_i)=1/2^8 (i=0,1,\ldots 255)$, then the entropy is $H(m)=8$ which corresponds to an ideal case. Practically, the information entropies of encrypted images are less compared to the ideal case. To design a good image encryption scheme, the entropy of encrypted image close to the ideal case is expected.

Information entropy for a few images is shown in Table 5 which is $H(m)=7.73$ and very close to the ideal value. This means a high permutation and substitution is achieved by the proposed algorithm and has a robust performance against the entropy attack.

Table 5. Entropy for different images.

| Images | Plain images | Cipher |
|---|---|---|
| Lena | 7.4451 | 7.7333 |
| Baboon | 7.1839 | 7.7289 |
| Water | 6.7754 | 7.7365 |
| Tiger | 7.2367 | 7.7315 |
| Beach | 6.6064 | 7.7339 |

### 3.2 Key sensitivity analysis

An ideal image cipher should be extremely sensitive with respect to the secret key used in the algorithm. Flipping of a single bit in the secret key, it should produce a widely different cipher image. This guarantees the security of a cryptosystem against brute-force attacks to some extent. Sensitivity with respect to a tiny change in the secret key for several images have been tested. One example for plain image '*Lena*' is discussed below:

a. Plain image (Figure 2(a)) is encrypted by using the secret key 'C6DA750B4C1F78D329 EA25E6B15CF9E47D21' and the resultant encrypted image is referred as image Figure 5(a).
b. The encrypted image (Figure 5(a)) is decrypted by making a slight modification in the original key '**B**6DA750B4C1F78D329EA25E6B15CF9E47D21' and the resultant decrypted image is referred as image Figure 5(b).
c. The encrypted image (Figure 5(a)) is decrypted by making a slight modification in the original key 'C6DA750B4C1F78D329EA25E6B15CF9E47D2**2**' and the resultant decrypted image is referred as image Figure 5(c).
d. The encrypted image (Figure 5(a)) is decrypted by making a slight modification in the original key 'C6DA750B4C1F78D329**F**A25E6B15CF9E47D21' and the resultant decrypted image is referred as image Figure 5(d).

With a small change in the key at any position, one is not able to recover the original image. It is not easy to compare decrypted images through the visual inspection. To compare decrypted images, correlation coefficient between encrypted images and various decrypted images was calculated and results are given in Table 6. The correlation coefficients are negligible.

Having the right pair of secret key is an important part while decrypting the image, as a similar secret key (with one bit change) will not retrieve the exact original image. Above example shows the effectiveness of the proposed technique as the decryption with a slightly different secret key does not reveal any information to an unauthorized person.

103

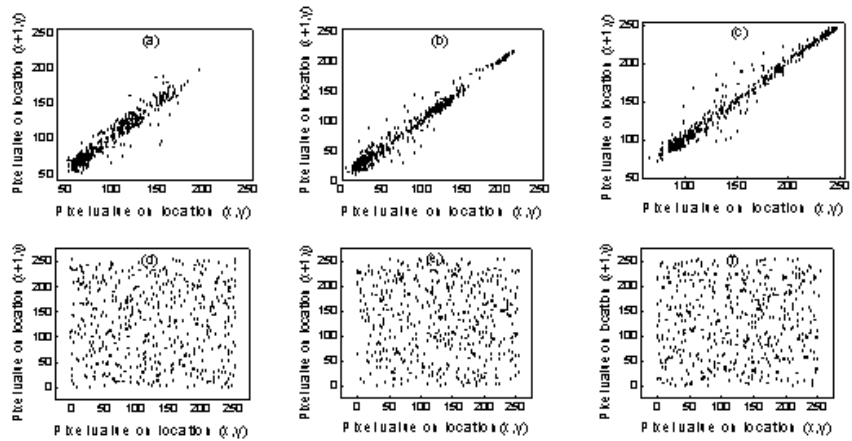

Figure 3 : Distributions of horizontally adjacent pixels of RGB components in the plain image '*Lena*' and its cipher image.

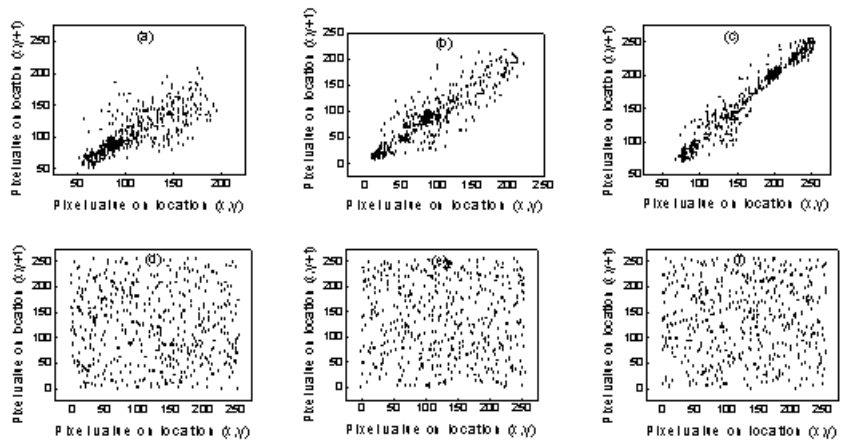

Figure 4 : Distributions of vertically adjacent pixels of RGB components in the plain image '*Lena*' and its cipher image.

Table 6. CR between RGB components of different decrypted images.

| Images | Correlation coefficient |
|---|---|
| Figure 5(a) and Figure 5(b) | $C_{RR}$= -0.00324, $C_{GG}$= -0.00124 , $C_{BB}$= 0.00273 |
| Figure 5(a) and Figure 5(c) | $C_{RR}$= -0.00119, $C_{GG}$= 0.00379 , $C_{BB}$= 0.00309 |
| Figure 5(a) and Figure 5(d) | $C_{RR}$= -0.00333, $C_{GG}$= 0.00356 , $C_{BB}$= 0.00304 |

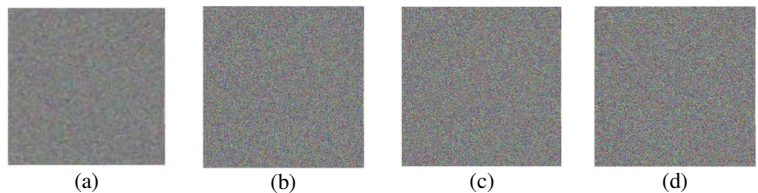

   (a)        (b)        (c)        (d)

Figure 5 : Decrypted images corresponding to plain image 'Lena' with slightly different secret keys.

International Journal of Network Security & Its Applications (IJNSA), Vol.4, No.2, March 2012

### 3.3 Differential attack

One minor change in the plain image causes large changes in the cipher image then differential analysis may become useless. Thus, much difference between encrypted forms is expected in order to keep high security. NPCR and UACI become two widely used security analyses in the image encryption community for differential attacks. NPCR concentrates on the absolute number of pixels which changes value in differential attacks while the UACI focuses on the averaged difference between two paired cipher images [17].

Suppose cipher images before and after one pixel change in a plaintext image are $c^1$ and $c^2$ respectively. The pixel value at grid (i,j) in $c^1$ and $c^2$ are denoted as $c^1(i,j)$ and $c^2(i,j)$ and a bipolar array D is defined by Equation (3). Then the NPCR and UACI can be mathematically defined by Equation (4) and (5) respectively

$$D(i,j) = \begin{cases} 0, & if\ C^1(i,j) = C^2(i,j) \\ 1, & if\ C^1(i,j) \neq C^2(i,j) \end{cases} \quad \ldots\ldots(3)$$

$$NPCR: N(C^1, C^2) = \frac{\sum_{i,j} D(i,j)}{T} X 100\% \quad \ldots\ldots(4)$$

$$UACI: U(C^1, C^2) = \frac{\sum_{i,j} |C^1(i,j) - C^2(i,j)|}{FxT} X 100\% \quad \ldots\ldots(5)$$

here symbol $T$ denotes the total number pixels in the cipher image, symbol $F$ denotes the largest supported pixel value compatible with the cipher image format and |.| denotes the absolute value function. Table 7 shows the values of NPCR (>99%) and UACI (≈33%) for each colour component image for four widely different nature images. Experimental results show the estimated expectations and variance of NPCR and UACI are very close to the theoretical values, which justify the validity of theoretical values. Hence the proposed encryption scheme is resistant against differential attacks.

Table 7. NPCR and UACI.

| Image | Dimension | NPCR of different colour components | | | | UACI of different components | | |
|---|---|---|---|---|---|---|---|---|
| | | Red | Green | Blue | Whole image | Red | Green | Blue |
| Lena | 512x512 | 99.1001 | 98.9746 | 98.9243 | 99.9966 | 33.2129 | 33.1642 | 33.1738 |
| Baboon | 200x200 | 99.4200 | 98.9350 | 98.8850 | 99.9925 | 33.2791 | 32.9219 | 32.9495 |
| Peppers | 200x200 | 99.3875 | 98.8350 | 98.8300 | 99.9900 | 33.2857 | 32.8767 | 33.0141 |
| Lion | 640x466 | 98.8910 | 98.8771 | 98.9875 | 99.9970 | 33.2543 | 32.9872 | 33.1372 |

### 3.4 Measurement of encryption quality

A very useful measure of the performance of the decryption procedure is the Peak Signal-to-Noise Ratio (PSNR). The PSNR of a given colour component is the ratio of the mean square difference of the component for the two images to the maximum mean square difference that can exist between any two images. It is expressed as a decibel value. The greater PSNR value (>30dB), the better the image quality recovered. For encrypted image, smaller value of PSNR is expected. Let P and P' being a plain image and cipher image respectively, the PSNR for the each color component (RGB) is defined as

$$MSE = \frac{\sum_{m=1}^{M} \sum_{n=1}^{N} [P(m,n) - P'(m,n)]^2}{MxN} \quad \ldots\ldots(6)$$

$$PSNR = 20*\log_{10}(255/sqrt(MSE)) \quad \ldots\ldots(7)$$

PNSR for the red, green and blue components of the cipher images with respect to their plain images for several images have been calculated. In all test cases, it was found to be small (<10dB). The computed results of peak-signal-to-noise ratio for various layers of image 'Lena' and its cipher image are given in Table 8. The low PSNR values reflect the difficulty in retrieving the plain image from the cipher image, without the knowledge of secret key.

Table 8. Peak Signal-to-Noise Ratio for different colour components.

| Images | PSNR (dB) | | |
|---|---|---|---|
| | Red | Green | Blue |
| Lena | 7.8703 | 8.5493 | 9.5924 |
| Baboon | 8.9919 | 9.6007 | 8.6377 |
| Beach | 6.1379 | 7.7897 | 9.2412 |
| Kids | 7.8596 | 8.4661 | 7.9083 |

### 3.5 Key space analysis

Secret key used in the image cipher should be neither too long nor too short. A larger secret key decreases the encryption speed and is not preferred for real time image transmission whereas a choice of smaller secret key results in an easy cryptanalysis. From the cryptography point of view, the size of the key space should not be smaller than $2^{100}$ to provide a high level of security [18,19]. Since the secret key is 144-bits long, the key space is about $2^{144}$ ($2.23 \times 10^{43}$). An image cipher with such a large key space is sufficient for resisting various brute-force attacks. It is possible to increase the number of bits for key in hardware implementation. However, by increasing the key length, volume of hardware is increased and consequently speed of the system is decreased.

### 3.6 Speed performance

Apart from the security consideration, encryption/decryption rate of the algorithm is also an important aspect for a good image cipher. Time taken by the proposed cipher to encrypt/decrypt various different sized colour images has been measured. The time analysis has been done on a personal computer with Intel core 2 duo 1.8Ghz processor and 1.5GB RAM. The results are summarized in Table 9, which clearly predicts an average encryption rate of proposed scheme is 450 KB/second.

Table 9. Encryption rate of proposed image cipher.

| Image size | Average |
|---|---|
| 200x200 (117 KB) | 0.22s |
| 512x512 (768 KB) | 1.76s |
| 640x480 (900 KB) | 2.04s |
| 800x600 (1.37 MB) | 3.24s |

## 4. CONCLUSIONS

A new non-chaotic digital image encryption scheme using a secret key of 144-bit size is proposed. Proposed algorithm uses both pixel substitution as well as pixel permutation process. In the substitution process, sub-block pixels value is modified which depends on used secret key as well as its adjacent previous and next sub-block. After partial encryption of each sub-block, used secret key is modified from recently encrypted sub-block. In permutation process, pixels position is reshuffled within sub-image by using key dependent generated magic square matrix. A feedback mechanism, used in substitution process, makes the encryption scheme robust and avoids differential attack. Proposed image cipher has high encryption rate, requires less computation and sensitive to used secret key. An extensive study of security and performance analysis of the proposed image encryption technique using various statistical analysis, key sensitivity analysis, differential analysis, key space analysis, speed performance, etc. have been carried out. Based on results of analysis, it is concluded that the proposed image encryption technique is perfectly suitable for the secure image storing and transmission.

**Narendra K Pareek** received M.Sc. from University of Rajasthan, Jaipur, India in 1986 and Ph.D. degree in Computer Science from M L Sukhadia University, Udaipur, India in 2005. He has worked as lecturer in the Department of Computer Science, Banasthali University, Banasthali, India for a period of two years. Presently, he is working as a Programmer at the University Computer Centre of the M L Sukhadia University, Udaipur since 1991 and has been teaching various courses of computer science to undergraduate and post graduate students. His research interests are in information security, chaotic cryptology, data compression, and information retrieval systems. He has published one book and more than twenty papers on chaotic cryptography and image encryption in referred 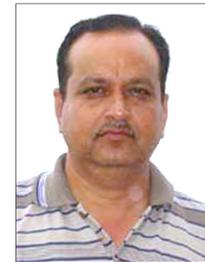 international journals/conferences. His research work has received more than 400 citations. His papers have also received MOST CITED PAPER AWARD for Image and Vision Computing journal and BEST PAPER AWARD in international conference (ICM2ST, 2011). He is also a senior member of Computer Society of India.